\documentclass{elsart}

\usepackage{natbib}

 \usepackage{graphics}
 \usepackage{graphicx}

\usepackage{amssymb}

\def\url#1{{\ttfamily\def\/{/\discretionary{}{}{}}#1}}

\begin{document}

\begin{frontmatter}
\title{r-Process Abundance Universality and Actinide Cosmochronology}
\author[nd,nao,int]{Kaori Otsuki},
\author[nd]{Grant J. Mathews},
\author[nao,int]{Toshitaka Kajino}
\address[nd]{Center for Astrophysics, 
University of Notre Dame, Notre Dame,IN46556-5670}
\address[nao]{Division of Theoretical Astrophysics, National Astronomical Observatory of Japan,Mitaka2-21-1, Tokyo181-8588 Japan}
\address[int]{Institute for Nuclear Theory, University of Washington, Seattle,
WA 98195}

\begin{abstract}
We review recent observational and theoretical results concerning 
the presence of actinide nuclei on the surfaces of old halo stars
and their use as an age determinant.
We present model calculations which show that 
the observed universality of abundances for
$56 < Z < 75$ elements in these stars 
does not necessarily imply a unique astrophysical site 
for the $r$-process.
Neither does it imply
a universality of abundances of nuclei outside of this range.
In particular, we show that a variety of astrophysical $r$-process models 
can be constructed
which reproduce the same observed universal $r$-process curve 
for $56 < Z < 75$
nuclei, yet have vastly different abundances for $Z \ge 75$ and possibly
$Z < 56$ as well.
This introduces an uncertainty into
the  use of the Th/Eu chronometer as a means to estimate 
the ages of the metal deficient stars.
We do find, however, that the U/Th ratio is a robust chronometer. 
This is because the initial production ratio of U to Th is 
almost independent of 
the astrophysical nucleosynthesis environment.
The largest remaining uncertainties in the U/Th initial production ratio
are due to  the input nuclear physics models.  

\end{abstract}

\end{frontmatter}

\section{Introduction}
Rapid neutron-capture (the $r$-process) is responsible for 
producing about half of the elements heavier than iron.  It is
believed to occur in an explosive stellar environment in which
the neutron capture time scale is much shorter than typical beta-decay lifetimes
near the line of stable nuclei.
The nuclear reaction flow can then proceed through  extremely neutron-rich unstable nuclei.
The fact that the heavy radioactive actinide nuclei, U and Th, are generated in the $r$-process
is of particular interest.
These nuclides have half lives [$t_{1/2}(^{238}$U)$= 4.47 \times 10^9~y$, 
$t_{1/2}(^{232}$Th)$= 1.40 \times 10^{10}~y$] which are comparable to the 
cosmic age.
These chronometers have taken on renewed recent attention 
as their absorption lines  have been identified \citep[e.g.][]{sneden,cayrel,
honda03a, honda03b}
in metal deficient stars.

The inferred abundances of Th and/or U 
can be used to estimate stellar ages.
Metal deficient stars are believed to be
the oldest stars in the Galaxy, and their surface abundances have probably
not changed (except for radioactive decay) since these stars were formed.
Moreover, their age can be regarded as the Galactic age and a lower limit to the cosmic age.
As a chronometer, this method is particularly appealing since
it avoids the usual Galactic chemical evolution model 
dependence \citep{meyerschramm}
associated with Solar-System radio cosmochronometry.
The Th or U on the surface of an old low-metallicity halo star were 
probably generated in a single nucleosynthesis event. 
Hence, the surface abundance of radioactive element $Y_{r}$
(r=Th or U) is given to a very good approximation by
\begin{equation}
Y_r (\Delta T)=Y_r(0)exp(-\Delta T/\tau_r)
\end{equation}
where $\Delta T$ is
time since nucleosynthesis and $\tau_r$ is the mean life of the r-element,
{\it i.e.} $\tau_r=t_{1/2}/ln 2$. 
For each radioactive element, one can solve Eq. (1) to find $\Delta T$.
It is best to utilize abundance ratios relative to an element with 
a nearby absorption feature (e.g. Eu).
Therefore, $\Delta T$ is 
usually given by 
\begin{equation}
\Delta T = 46.7 (\log ({\rm Th/Eu})_{0}-\log ({\rm Th/Eu})_{\rm T})~~,
\end{equation}
\begin{equation}
\Delta T= 21.8 (\log({\rm U/Th})_{0} - \log({\rm U/Th})_{\rm T})~~,
\end{equation}
in units of Gyr, where the index $0$ denotes the initial production ratio, 
while the index $T$ 
refers to the presently observed value.
(In these equations we denote $Y_{Th}=Th$, $Y_{Eu}=Eu$, and $Y_U=U$.)
The only uncertainties are, therefore, those in the determination of the
present stellar abundances themselves, and those due to uncertainties in
model estimates of the initial production ratios.
[That is, as long as the produced actinide nuclei have not passed through
stellar CNO burning where they might be destroyed by photo-induced fission
\citep{mmd89}.]
 
In view of the significance of this independent measure of the Galactic age,
it becomes important to scrutinize and quantify these remaining 
uncertainties as much as possible.  In this paper we are primarily concerned 
with the astrophysical uncertainties in the initial production abundances.
We show that there is indeed considerable uncertainty in using only a single
Th or U radiochronometer, even when the universally observed 
Solar-System $r$-process abundances are well
reproduced for lower-mass nuclei.  We also establish
that the Th/U chronometer is quite robust and somewhat independent 
of astrophysical model uncertainties.

For some time \citep[cf.][]{truran,mathews90} 
r-process elements in metal deficient stars have been interpreted as evidence
for a universal $r$-process abundance distribution in 
operation in the early Galaxy.
In particular, more recent observations 
\citep{sneden,Sned98,sneden2,johnson}
all show similar abundance distributions for $Z > 56$ elements.
This feature is often referred to as the ``universality'' of the $r$-process.
Hence, it is generally believed that, at least for $Z > 56$ elements,
the astrophysical site and associated yields of 
r-process nucleosynthesis are unique.
Based upon this assumption, \citet{sneden} estimated the ages of several
stars using the ratio of Th/Eu at the time of 
formation of the Solar System as the
initial production ratio [even though Solar-System material has 
experienced multiple supernovae 
before its formation and has experienced Th decay]. 
The analyses of these stars 
all indicated similar present ages of about 14 Gyr $\pm$ 4.

There are, however, some reasons to question the assumption of 
a universal $r$-process
abundance curve.  The material out of which these metal poor stars 
were formed is
likely to have experienced only
one or two supernovae before incorporation into the star.  
Depending upon which particular progenitor supernova was in operation, 
there might be substantially different abundance distribution
curves for these stars, compared to the ensemble average represented in
Solar-System material \citep[cf.][]{Ishi99}.
Moreover, quite recently, \citet{cayrel} 
have reported the observation of 
peculiar r-process elements in the metal deficient star CS31082-001.
This is the first star for which a uranium line was also detected.
This star has the strange feature, however,  
that the ratio of Th to Eu is greater than 
that of Solar material. This would imply (on the bases of Th/Eu)
that this star is younger than the Sun. 
This seems unlikely in view of its low metallicity, [Fe/H] $\sim -2.9$. 
Furthermore, the Th/Eu age is in contradiction with the U/Th age for this star
which is 12 $\pm$ 3 Gyr.
Further evidence of Th/Eu uncertainty is apparent in the recent data of
\citet{honda02} and \citet{honda03a,honda03b}.
They have reported on r-process element abundance distributions 
in two other metal-deficient stars.
These stars also show different abundance distribution patterns 
for $Z \ge 75$.

Even the stars studied in Sneden et al.(1996) and Sneden et al.(2000)
may show deviations from 
a universal $r$-process distribution for lighter Z $<$ 56 nuclei.
For example Sneden et al.(2000) noted divergence from the solar r-process for a few elements in CS22892-052.
\citet{cowan} also noted some divergence in the star 
BD +17 3248, but those data are somewhat uncertain.  
This, together with meteoritic evidence, has
been taken as an indication \citep{qian} that two different
$r$-process environments could be in operation.
All in all, the observational data  seem to indicate that the universality of 
r-process elements 
may be broken for $Z \ge 75$ elements and possibly Z $\leq$ 56 as well.

In this paper, we wish to clarify the astrophysical model dependence
for these stellar r-process chronometers.
Several previous investigations \citep{gor,wanajo02,sha,q02,otsuki03}
of the uncertainties in actinide chronometers can be found in the literature.
The paper by \citet{gor}
and \citet{sha} 
were primarily directed toward understanding the nuclear uncertainties.
For example, \citet{gor} 
considered 32 different models for $r$-process actinide production
based upon different nuclear physics input for fission barriers, 
nuclear masses, beta-decay rates, etc.
In \citet{sha}, it was found that the Pb abundance was useful to 
constrain nuclear models.
These calculations, however, were made in the context of schematic ``canonical event'' models which were constrained to reproduce Solar-System 
$r$-process abundances.
Hence, universality for all elements up to Z=82 was imposed.
In another related work, \citet{wanajo02} have attempted to clarify 
the age of CS31082-001 in the context of a specific ``neutrino-driven wind''
model, but with only electron fraction and the 
outer boundary temperature as free
parameters.
\citet{q02} even shows on the basis of the astrophysical observations that 
the production of the elements with $A>130$ is not robust.
All of these works have demonstrated some uncertainties 
in the Th/Eu chronometer, while better results are obtained using the U/
Th chronometer.

The present work differs from the previous works in two important ways:
1) Unlike \citet{gor} and \citet{sha}, our focus is on the astrophysical 
rather than nuclear uncertainties;
2) Rather than to focus on a particular wind model as in 
\citet{wanajo02} and \citet{otsuki03}, 
we consider wide range of plausible astrophysical models and
parameters all of which reproduce the universality in the abundances for 
56 $<$ Z $<$ 75 noted in observations.  
We also explicitly considerneutrino interaction effects.

From the observations and our theoretical calculations, we conclude that the
astrophysical site for r-process nucleosynthesis is probably not unique, 
even though there is an observed universality of abundances.
That is, a variety of astrophysical
models can be constructed which reproduce the same 
apparent universal abundance curve
for elements with $56 < Z < 75$, but which give large variations in 
abundances for elements with $Z \ge 75$ and/or $Z \le 56$.
Hence, it is dangerous to use the Th/Eu chronometer to estimate 
the age of metal deficient stars.
Moreover, the production ratio of the Th/Eu
chronometer is strongly dependent upon  the nucleosynthesis environment.
On the other hand, we demonstrate that
the U/Th production ratio is almost independent 
of the nucleosynthesis environment, confirming that this ratio is robust as a  chronometer.

In what follows, we first review in more detail 
the recent observational results for 
metal deficient halo stars in section 2.
Results of theoretical calculations are shown in section 3.
In section 4, we will summarize the viability of the nuclear cosmochronometers.

\section{Observational data}
As noted above, it has been shown \citep{sneden,Sned98,
sneden2} that 
the star CS22892-052 confirmed the previously noted fact 
\citep{spite, truran,
gilroy} 
that r-enhanced stars show an abundance distribution which is
similar to that of Solar r-process material for Z$>$56 elements 

Several lighter elements (Z$\le$56), which would also  be generated in 
the r-process, 
were also detected in this star.
These elements, however,  showed a  different abundance distribution from 
that of Solar material.
The agreement between the heavier elements (Z$>$56) and 
scaled Solar-System abundances
has been reported for other metal deficient stars
[e.g. HD115444 \citep{westin}, HD126238, HD186478,
HD108577 \citep{johnson}].
In addition, there is also evidence (Sneden et al. 2002; Aoki et al. 2003) that
the europium isotope ratios in several r-process enhanced stars are
consistent with the Solar r-process isotope ratio.  This further
enhances the evidence of a single universal abundance curve for $Z > 56$
elements.
Hence, at least for Z$>$56 elements, 
it is generally believed that astrophysical site for r-process nucleosynthesis
is unique.
If one can therefore assume that 
the Solar-System and metal-deficient stars have 
the same abundance distribution
when they form,  then one can use the Solar-System abundances 
as the initial production ratio.
This approach has the problems that the material in the Solar System 
represents an average of the ejecta from many supernovae,
and the Th and U present when the Solar System formed will have
decayed from its initial production ratio.  
Nevertheless, this method is still
useful as a means to estimate an upper limit on the age of these stars.
Several works have estimated
 the ages of those metal deficient stars
using a scaled Solar Th/Eu value \citep{sneden,Sned98,sneden2,johnson}.

However, as noted above, 
\citet{cayrel} have reported on a uranium line for 
the star CS31082-001.
Although their abundance distribution shows 
a similar pattern to the Solar System
for 56 $<$ Z $<$ 75 elements, there are large differences for Z $\ge$ 75
elements.
In addition to 
the fact that the Th/Eu abundance ratio is larger than in the 
Sun, this star has less Pb than in Solar material \citep{hill}.

These two facts are strange because U and Th decay into Pb.
As noted above, 
these facts would  indicate that this star is much younger than the Sun 
in spite of its lower metallicity.
Moreover, this age is, however, in conflict with the U/Th age fot this star.

Furthermore, \citet{honda02} and \citet{honda03a, honda03b} have 
reported on r-process elements
in several more metal-deficient stars which were observed using
the SUBARU/High Dispersion Spectrometer(HDS).
Of particular interest are the stars 
HD6268 and CS30306-132 \citep{honda02, honda03a, honda03b}.
These two stars show a  large Th/Eu ratio which is  comparable to
that of Solar material
in spite of their lower metallicity.

In summary, the observations all seem to indicate the following:
1) For 56$<$Z$<$75 elements: the r-process abundance distribution 
appears to obey a universal curve;
2) For Z$\ge$75 and some Z$<$56 elements: the r-process abundance 
distribution may not always coincide
with a universal abundance curve.

All of this implies that the r-process nucleosynthesis environment 
may not be unique even though a universal abundance curve is produced
for 56 $<$ Z $<$ 75. 
In the next section we seek to explain these trends in the context of
models for the $r$-process.

\section{Theoretical calculation}
We  wish to clarify the dependence of the abundance distribution 
on the astrophysical environment.
Although it is generally believed 
that the astrophysical site for the r-process is an explosive event,
the precise environment has not yet been unambiguously identified.
Type II SNe or neutron-star mergers have  been proposed 
as two possible sites (e.g. Mathews \& Cowan 1990)
with the neutrino-energized wind above the proto-neutron
star \citep{Woos94} being the currently most popular paradigm 
[see however Qian and Woosely (1996), Cardall and Fuller (1997) 
and Meyer et al.(1998)]. 
In the following, we base our schematic parameter study on
two dynamical models. 
One is the 
neutrino-energized wind
model \citep{Woos94,Otsu00}.
The other is an exponential model. 
We believe that
these parameter studies should 
approximate the gamut of possible r-process models 
such as might also occur, for example, in
neutron-star mergers or prompt supernova explosions.
  
\subsection{Hydrodynamic wind model}
For the wind model,
we use a spherical steady-state flow apploximation for the neutrino-driven wind
(cf. Qian \& Woosley 1996; Qian 2000; Takahashi \& Janka 1997; 
Otsuki et al. 2000).
The hydrodynamic flow is deduced from
the following non-relativistic  equations:
\begin{equation}
4 \pi r^2 \rho v = \dot{M}~~,
\end{equation} 
\begin{equation}
\frac{v^2}{2} - \frac{M G}{r} + N_A k T s_{rad} =E~~,
\label{eeq}
\end{equation}
\begin{equation}
s_{rad}(\leq s)\sim s_{rad}^{(0)}=\frac{11 \pi ^2}{45 \rho N_A}\
(\frac{k T}{\hbar c})^3~~,
\end{equation}
where $\dot M$ is the rate at which matter is ejected by neutrino heating 
from the
surface of the proto-neutron star, and  
$k$ is the Boltzmann constant.
In equation (\ref{eeq}), $M=1.4M_{\odot}$ is the protoneutron star mass,
the total energy $E$ is fixed by
a boundary condition on the asymptotic temperature $T_b$ 
\begin{equation}
E = N_A  s_{rad} k T_b~.
\end{equation}

For simplicity, in the present work we utilize an adiabatic 
(constant entropy) wind rather than to compute the neutrino heating 
explicitly \citep[e.g. as in][]{Otsu00,wanajo}.
This is adequate for our purpose which is to sketch the
dependence of the abundances on the
nucleosynthesis environment.
Nevertheless explicit charged and neutral-current 
neutrino-nucleus interaction effects on abundances are included 
as described below.

Using this model, we can calculate time profiles of temperature, matter,
and the neutrino density 
for material flowing away from the proto-neutron star. 
These are then input into a nucleosynthesis network
where the neutron density and abundance yields
are computed.
The coordinate r was chosen such that when $T_9=9.0$, $r=14$ km.
That way all trajectories experience the same initial neutrino luminosity.

\subsection{Exponential models}
There are several theoretical calculations for r-process nucleosynthesis
with a longer timescale than that of the wind models, e.g. in
prompt supernovae.
To model this we have used the following parameterized
temperature and density profile.
\begin{equation}
T_9=9.0 exp(-t/t_{exp})+0.6 , 
\end{equation}
\begin{equation}
\rho = 3.3 \times S/T^3.
\end{equation}
Here, S is the entropy and $t_{exp}$ is the dynamical timescale.

\subsection{Nucleosynthesis}

Our nucleosynthesis code is based upon the dynamical network calculation
described in Meyer et al. (1992) and Woosley et al. (1994), but with 
several important modifications.
For one, an extensive light-element neutron-rich 
reaction network has been added
as described in Teresawa et al. (2001) and Orito et al. (1997).  This 
light-element network can be important for high-entropy 
short dynamical timescale environments.
We also compute the alpha-rich freezeout and the $r$-process in a single
network rather than to split this calculation into two parts as
was done in Woosley et al. (1994).
Some results of our calculations for various parameters are shown on Fig. 1.
The calculations shown on this figure 
are for a constant neutrino luminosity
of $L_{\nu} = 10^{51}$ erg sec$^{-1}$. 
In Fig.2, we show the dependence on neutrino luminosity.
We have included effects of
neutrino-nucleus interactions on the nucleosynthesis yields.
For more details of this code, 
see Terasawa et al. (2001) and reference there in.

\subsection{Results}

A sample of model parameters studied in the present work  are summarized in
Table 1 and Figures 1-3.  As noted below, most of these models reproduce the
observed universality of elemental abundances for  $ 56 < Z < 70$.  For
illustration
we also show the implied europium isotope fraction 
\\
$fr(^{151}Eu) \equiv
^{151}Eu/(^{151}Eu
+ ^{153}Eu)$.  Most of these models are also consistent with the
observed isotope fraction
of  $fr(^{151}Eu) = 0.5 \pm 0.1$ (Sneden et al. 2002; Aoki et al. 2003).

\subsubsection{Wind model}

Figures 1a-d and 2 
illustrate the dependence of the nucleosynthesis yields on 
various parameters of the wind models.
Note, that all of these models are  constrained to reproduce 
the observed universal abundances for $56 < Z <75$ as shown 
by the dots on Figure 1a.
We have not included the decay back to Pb from long-lived U and Th.
Here we only require that the initial production of
Pb is less than the observed Pb to account for this decay back.
Ultimately, a quantitative account of Pb is important (Schatz et al. 2002) 
for constraining the nuclear physics input to the r-process and
will be considered in a subsequent work.

Figure 1a shows the effects of entropy.
Changing the entropy essentially alters  
the neutron-to-seed ratio as the $r$-process begins.
Note, that all of these models show an almost identical
abundance distribution for $56 <Z< 75$ elements.
Clearly, the observed universality of the r-process is maintained
for these elements.
On the other hand, elements with $Z \ge 75$ show much different distributions.
Obviously, the Th/Eu ratios would be substantially different for models
with different entropy, even though the universal abundance
distribution for $Z \le 75$ elements is obtained.

We have also studied effects of altering 
the electron fraction (Fig. 1d) and the dynamical time scale (Fig. 1c).
These parameters similarly affect the 
neutron-to-seed ratio at the onset of the $r$-process,
and produce nearly equivalent results to those shown on Figure 1a,
i.e. the  abundance distribution
for $Z < 75$ elements is unaffected, while elements with $Z \ge 75$
are significantly altered.
Note that varying $Y_e$ has the smallest effect on the distribution,
and hence, does not by itself give a fair representation of the model
uncertainties.

The asymptotic temperature $T_b$ at which neutron captures freeze out
also affects the abundance distribution as shown in Figure 1b.  
In this case,
the freezeout temperature essentially determines the $r$-process path
at the termination of the $r$-process.  
A discussion of this dependence on asymptotic temperature also appears 
in Terasawa et al. (2002) and Wanajo et al. (2002).
Note, that in all three of the models chosen, the abundances
between  $60 \le Z \le 75$ are consistent with the universal curve.
In this case, however, both elements with $Z < 60$ and $Z > 75$
are substantially different. 

In Figure 2 we also show the dependence upon neutrino luminosity.
Neutrino-induced beta-decay can also affect the abundance distribution.
The main effect of neutrino luminosity in these models is to provide 
neutrino-induced neutron emission.
For $60<Z<70$, this helps to smooth the abundance distribution, 
especially just above the r-process peaks.
Neutrino interactions however can also significantly 
affect the heavy $Z>70$ abundances.
This is because neutrino capture on neutrons can increase $Y_e$
and decrease the neutron to seed ratio. 

\subsubsection{Exponential model}
Results for the exponential models are shown in Fig. 3.
If the dynamical timescale becomes longer than around 1.5 seconds, 
the yields begin to differ from the universal pattern. 
The longest dynamical timescale case shows a different pattern even 
for $ 56 < Z < 70$ elements. 
It also shows a Eu isotopic ratio which is inconsistent with the observed
values.
These differences relate to the fact that near freezeout neutron captures are
in competition with beta decay in these models.
Hence, the r-process path at neutron capture freeze out is different.
As discussed below, if we could resolve nuclear physics uncertainties, 
the observed abundance distribution might discriminate between short and long 
timescale astrophysical sites for the r-process. 

\subsection{Analysis of universality}
Both observations and our theoretical calculations indicate 
that the observed universality of the $r$-process does not imply a unique
astrophysical site for r-process nucleosynthesis.
That is, a variety of conditions can lead to the same observed
universal abundance distribution in the $56 < Z < 75 $ region.
 The reason for the universality of
the $60 \le Z \le 75$ elements can be traced to the fact
that there are no significant nuclear shell closures 
along the $r$-process path for the progenitor nuclides of
these elements.
The $r$-process path corresponding to $(n, \gamma ) \rightleftharpoons  
(\gamma,n)$ equilibrium essentially flows along a fixed neutron separation
energy given approximately as
\begin{equation}
S_n = \frac{T_9}{5.040} \times \{34.075-log(Y_n \times \rho) + 
1.5 \times log(T_9)\} {\rm ~MeV},
\end{equation} 
where $\rho$ is the mass density of material in units of g cm$^{-3}$,
$T_9$ is the temperature in units of $10^9$ K, and 
$Y_n$ is the neutron number fraction.  In the models shown on Figures 1a-d,
there are no waiting points for the progenitors of $56 < Z < 75$ elements. 
The progenitor
nuclei have larger neutron capture cross sections.
Hence $(n,\gamma) \rightleftharpoons
(\gamma,n)$ equilibrium is maintained longer for these elements.
Moreover, for a variety of astrophysical conditions 
the r-process path at freezeout is so far from stability 
that the final abundance distribution is dominated by 
the effects of beta-delayed neutron emission which leads 
to the same universal curve.

\section{Cosmochronometry} 
\subsection{Th/Eu chronometer}
Our theoretical calculations show that the ratio of Th to 
Eu (or to any other 
stable r-process nuclei for 56$<$Z$<$ 75) is strongly dependent upon
the nucleosynthesis environment.
Table 1 summarizes Th/Eu ratios calculated in different environments most of 
which reproduce the observed universality and have initial 
Th/Eu ratios greater than that presently observed in CS22892-052.
These Th/Eu ratios vary by almost two orders of magnitude rendering
an uncertainty in the inferred ages of at least a factor of 2.

To estimate the age of a metal deficient star using 
the Th/Eu chronometer,
one must therefore unambiguously identify the astrophysical 
site at which they were generated.
This is particularly true for the oldest metal-poor stars whose
initial composition is 
likely to have been enriched by only a single supernova event.
The details of that particular supernova event are likely to
be different from one metal poor star to another.
Therefore, it is dangerous to estimate the ages of metal deficient stars
using the Th/Eu ratio.
One cannot predict the initial Th/Eu 
production ratio without identifying
its nucleosynthesis environment.
We note, however, that the sensitivity of light-element abundances and
Pb to parameters (e.g. Fig.~1d)
might be used to help identify this environment.

\subsection{U/Th chronometer}
On the other hand, because Th and U are neighboring actinide nuclei, 
and because both
are the result of an extensive alpha decay chain
(Cowan, Thieleman, and Truran 1991), one would anticipate that
some of the above model dependence for this chronometric pair
might be removed.  We have studied various models of different entropy,
$Y_e$, and freezeout temperature under the constraint that
the abundances of 56 $<$ Z $ <$ 75 elements be
consistent with the observational results,
and also that the initial production ratio of Th and U to Eu should be 
 significantly larger than the presently
observed value in CS22892-052 of $log(Th/Eu) = -0.66 \pm 0.08$  (Cowan et al
1999).

Table 1 summarizes various models.
Three of the models shown in Figures 1-3 are ruled out by these criteria.
The ratio of $^{238}$U/$^{232}$Th for each of these models 
is given.
It is particularly noteworthy that all
models yield almost the same value of the U/Th ratio even though the Th/Eu 
ratios are vastly different.
These representative models all lead to a production 
ratio of $U/Th=0.40 \pm 0.05 ~(1 \sigma)$.
Thus,
our calculations indicate that U and Th are nearly
always generated in the same ratio
as long as enough Th and U are generated 
to be consistent with  the observed abundances.
Since the stellar age only depends logarithmically of the production ratio 
by Eq. 1, 
this is sufficiently accurate to determine the stellar age to  $\pm  14\%$
or about $\pm 2$ Gyr.
 
Hence, the U to Th production ratio does not particularly
depend upon  details of the 
nucleosynthesis environment.
Using this chronometer, one can, therefore,  
estimate the age of metal deficient stars 
without much astrophysical uncertainty.
Unfortunately, the U to Th ratio does, however,  depend upon 
nuclear theoretical models [e.g. \citet{gor}], 
i.e. mass formulae, beta-decay rates, alpha-decay rates, and 
particularly $\beta$ -delayed fission and neutron emission \citep{meye89}.
We are presently conducting a separate study of the dependence of 
this ratio on the input nuclear physics.

\begin{figure}
\begin{center}
\includegraphics[width=14cm]{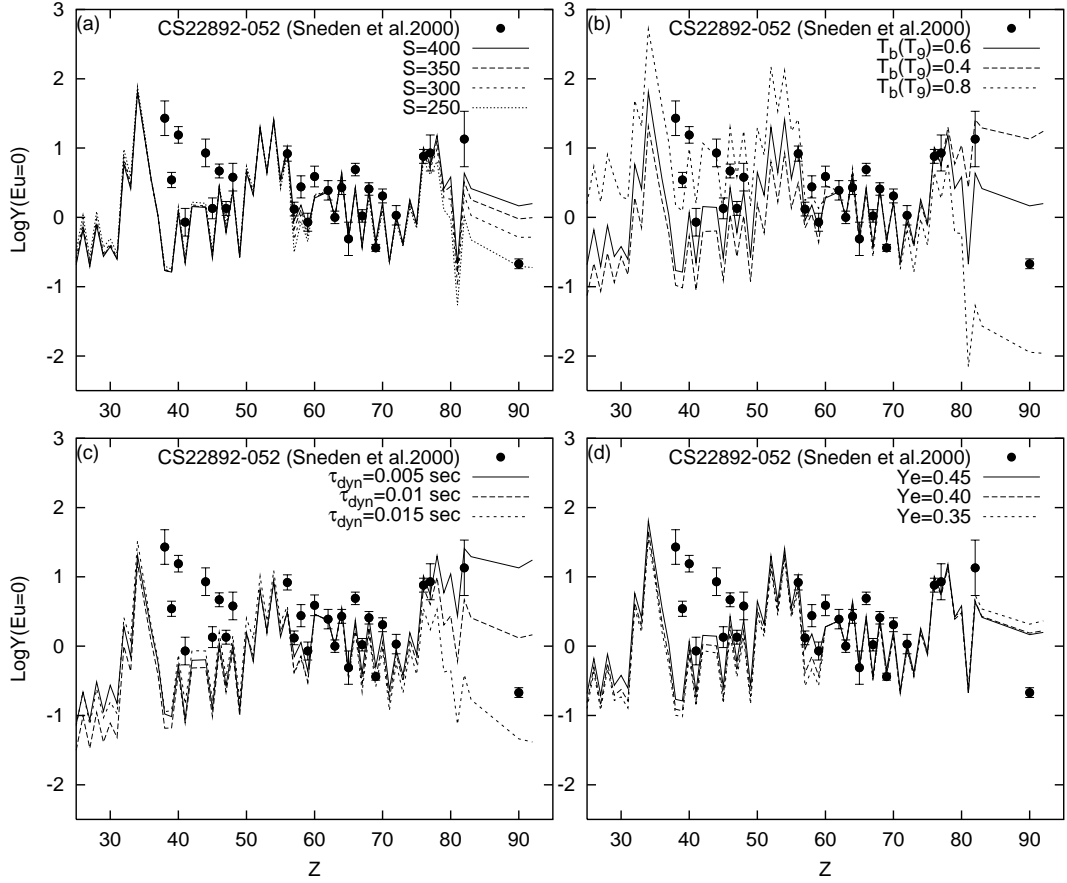}
\end{center}
\caption{Dependence of the nucleosynthesis yields 
upon various parameters of the astrophysical environment (see Table 1).
Closed circles show observed elemental abundances in CS22892-052
\citep{sneden2}.
In (a) 
the entropy per baryon for each model is $S/k=$ 400, 350, 300, and 
250 as labeled, 
and the dynamical time scale is fixed at $0.005$ sec.
In (b) the asymptotic temperature is 0.4, 0.6, 0.8
in units of $T_9$ as labeled.
In (c) dynamical time scales of $0.015, 0.010$, and $0.005$ sec
are shown for fixed S=400 and $Y_e$=0.45.
In (d) the electron fraction is varied.
(See Table 1 for details.) \label{fig1}}
\end{figure}

\begin{figure}
\begin{center}
\includegraphics[width=7cm]{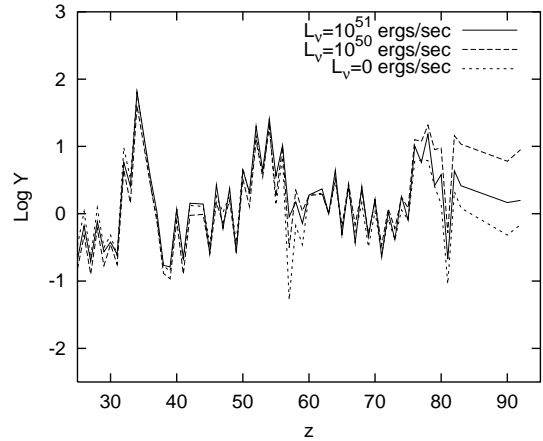}
\end{center}
\caption{Dependence of yields upon neutrino luminosity. }
\end{figure}

\begin{figure}
\begin{center}
\includegraphics[width=7cm]{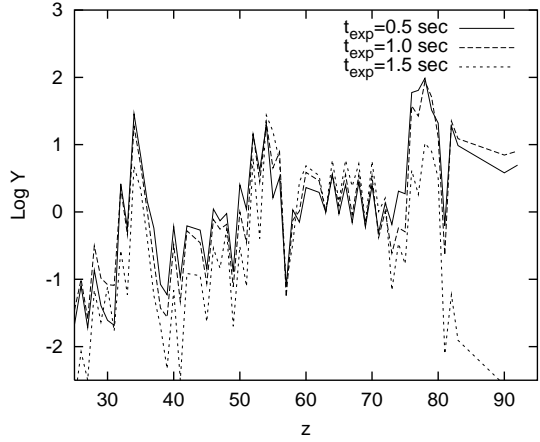}
\end{center}
\caption{Dependence of the yields upon timescale. 
If the timescale becomes longer than 1.5 second, the universal pattern is broken.}
\end{figure}

\begin{table}
\begin{center}
\caption{U/Th, Th/Eu, and parameters for various models shown in Figures
 1 - 3}
\begin{tabular}{lllllllll}
\hline
 S/k & ${\rm T_b (T_9)}$ & $\tau_{\rm dyn} (sec)$ & $Y_e$ & U/Th 
&Th/Eu&fr$({\rm Eu}^{151})$ &$L_{\nu}$ ergs/sec & Figure\\
\hline
400 & 0.4&$0.005$ & 0.45 & 0.51&13.5&0.53 &$10^{51}$&1(b),1(c)\\
400 & 0.6 & $0.005$ & 0.45 & 0.38&1.5&0.51&$10^{51}$&1(a),1(b),1(d),2\\
350 & 0.6 & $0.005$ & 0.45 & 0.37&0.95&0.52&$10^{51}$&1(a)\\
300 & 0.6 &  $0.005$ & 0.45 & 0.37&0.51&0.52&$10^{51}$&1(a)\\
250 & 0.6 & $0.005$ & 0.45 &0.36 &0.20 &0.52& $10^{51}$&1(a)\\
400 & 0.8 & $0.005$ & 0.45 &0.26&$0.0011$&0.47&$10^{51}$&1(b)\\
400 & 0.4 & $0.01$ & 0.45 & 0.40&1.3&0.54&$10^{51}$& 1(c)\\
400 & 0.4 & $0.015$ &0.45 & 0.27&0.045 &0.54 & $10^{51}$&1(c)\\
400 & 0.6 & $0.005$ & 0.35 & 0.39 & 1.5&0.52&$10^{51}$&1(d)\\
400 & 0.6 & $0.005$ & 0.40 &0.40 & 0.21&0.52 &$10^{51}$& 1(d)\\
400 & 0.6 & $0.005$ & 0.45 & 0.58 & 6.0 & 0.49 &$10^{50}$&2\\
250 & 0.6 & $ 0.005$& 0.45 & 0.54 & 0.48 & 0.49 &$0 $&2\\
400 & 0.6 & $0.5$ & 0.3 & 0.51 & 3.8 & 0.47 & 0 & 3 \\
400 & 0.6 & $1.0$ & 0.1 & 0.43 & 7.0 & 0.55 & 0 & 3 \\
400 & 0.6 & $1.5$ & 0.1 & 0.50 & $0.0027$ & 0.35 & 0 & 3 \\
\hline

\end{tabular}
\end{center}

\end{table}

\section{Conclusion and discussion}
Our theoretical calculations indicate that
the coincidence of the observed abundance distribution for $56<Z<75$ 
elements with the Solar r-process abundances does not necessarily  mean  that all $r$-process events occur
in the same universal environment.
Moreover, different abundance distributions for $Z \ge 75 $ and $Z \le 56$
elements are produced even when the universal $56 < Z < 75 $ abundances are
reproduced.
Hence, it is not possible to predict the initial production ratio of Th to Eu 
in each metal deficient star
without somehow independently identifying each production environment.
We therefore conclude that there is substantial uncertainty
when using the Th/Eu chronometer to estimate 
the ages of metal deficient stars.

We note that the reason for the uncertainty is that in realistic (e.g. SN wind)
r-process models, one tends to run out of neutrons as actinide 
elements are being formed.
It might be possible, however, that an r-process environment
could occur in 
which there were abundant neutrons present as the actinide nuclei are 
synthesized.
In this case, the r-process could proceed to the fission termination 
and recycling.
If most stars experienced this, then the uncertainty in the Th/Eu chronometer
would be reduced and
the CS31082-001 star would be an interesting exception.
Nevertheless, since it is difficult to produce fission recycling in realistic
(e.g. wind) models, we caution against this assumption.
Another possibility to reduce the Th/Eu uncertainty is
that a combination of light-element (Z $\le$ 56)
abundances and Pb-region abundances \citep{sha}
might help to identify the environment.
We also note that variations of the light-element abundances are quite natural
in neutrino-driven wind models even when a universality of heavier elements
is imposed.
Hence, one may not need to propose two different $r$-process sites 
\citep{qian} to account for the variations of 
light elements relative to the
universal heavy nuclei. Although two sites are probably still necessary to 
explain the meteoritic evidence.
To some extent, however, the light-element abundances may simply 
reflect the star-to-star
variations in a single neutrino-driven wind scenario for the $r$-process.

On the other hand, we find that the U/Th chronometer is a 
robust means to estimate the
cosmic age.
From our theoretical calculations,
we have shown that the U/Th initial production ratio is almost independent of
the nucleosynthesis environment as suggested by \citet{gor} and
\citet{wanajo02}.
Indeed, the present work vindicates the conclusions of those papers for wider
range of astrophysical environments.
Hence, using this chronometer,
one can estimate the age of metal deficient stars
with little correction for the astrophysical uncertainties.
Unfortunately, some uncertainty due to 
the nuclear physics still remains \citep[e.g.][]{sha}.
Further quantitative studies 
of the uncertainties due to various nuclear physics
input such as the mass formula, beta-decay rates, alpha-decay rates, 
$\beta$-delayed fission and neutron emission
are strongly desired. Efforts along this line are currently underway.

Regarding the universal abundance curve, we have shown that 
there is a dependence of the $56<Z<70$ abundances on the dynamical time scale.
We also note that there is a dependence of these abundances on 
the detailed nuclear physics models.
The hope is that with sufficiently precise observed abundances one might be
able to both identify the true dynamical timescale
and the correct nuclear physics models for the r-process.
More accurate observational data are likely to be available in the future
due to the development of
large aperture telescopes and high resolution spectroscopy. 
Clearly, more observations of $r$-process abundances in metal poor stars
are desired.

This work has been supported in part by Grants-in-Aid for
Scientific Research (10640236, 10044103, 11127220, 12047233)
of the Ministry of Education, Science, Sports, and Culture of Japan.
Work at the University of Notre Dame was supported
by the U.S. Department of Energy under
Nuclear Theory Grant DE-FG02-95-ER40934.
One of the authors (KO) wishes to acknowledge support
from the Japanese Society for the Promotion of Science.
KO and TK also thank the Institute for Nuclear Theory at the University of
Washington for their hospitality and the Department of Energy for their
partial support during the completion of this work.


\end{document}